\documentclass[10pt,journal,compsoc]{IEEEtran}
\usepackage{amsmath}
\usepackage{amssymb}
\usepackage[english]{babel}
\usepackage{slashbox}
\usepackage{times}
\usepackage{epsfig}
\usepackage{graphicx}
\usepackage{diagbox}
\usepackage{multirow}
\usepackage{color}
\usepackage{multirow}
\usepackage{booktabs}
\usepackage{verbatim}
\usepackage{wrapfig}
\usepackage{graphicx}
\usepackage{comment}
\usepackage{amsthm}

\usepackage[justification=centering]{caption}
\newtheorem{definition}{Definition}

\def\etal{\emph{et al}.~}

\def\ie{\emph{i.e}.~}

\ifCLASSOPTIONcompsoc
  \usepackage[nocompress]{cite}
\else
  \usepackage{cite}
\fi
\ifCLASSINFOpdf
\else
\fi
\begin{document}
\title{Geometry-guided Dense Perspective Network for Speech-Driven Facial Animation}

\author{Jingying Liu$^{\dagger}$, Binyuan Hui$^{\dagger}$, Kun Li$^{*}$,~\IEEEmembership{Member,~IEEE,}~Yunke Liu, ~Yu-Kun Lai,~\IEEEmembership{Member,~IEEE,}~Yuxiang Zhang, Yebin Liu,~\IEEEmembership{Member,~IEEE,} and Jingyu Yang,~\IEEEmembership{Senior Member,~IEEE}\\
         % <-this % stops a space

%E-mail: lik@tju.edu.cn
\IEEEcompsocitemizethanks{
\IEEEcompsocthanksitem  $\dagger$ Equal contribution.
\IEEEcompsocthanksitem $^{*}$ Corresponding author: Kun Li (Email: lik@tju.edu.cn)
\IEEEcompsocthanksitem Jingying Liu, Binyuan Hui, Kun Li and Yunke Liu are with the College of Intelligence and Computing, Tianjin University,
Tianjin 300350, China.

\IEEEcompsocthanksitem Yu-Kun Lai is with the School of Computer Science and Informatics, Cardiff University, Cardiff CF24 3AA, United Kingdom.

\IEEEcompsocthanksitem Yuxiang Zhang and Yebin Liu are with the Department of Automation, Tsinghua University, Beijing 10084, China.

\IEEEcompsocthanksitem Jingyu Yang is with the School of Electrical and Information Engineering, Tianjin
University, Tianjin 300072, China.\protect\\
}}


\markboth{IEEE Transactions on Visualization and Computer Graphics}%
{Jingying Liu \MakeLowercase{\textit{et al.}}: Geometry-guided Dense Perspective Network for Speech-Driven Facial Animation}



\IEEEtitleabstractindextext{%
\begin{abstract}
Realistic speech-driven 3D facial animation is a challenging problem due to the complex relationship between speech and face. In this paper, we propose a deep architecture, called \emph{Geometry-guided Dense Perspective Network (GDPnet)}, to achieve speaker-independent realistic 3D facial animation. The encoder is designed with dense connections to strengthen feature propagation and encourage the re-use of audio features, and the decoder is integrated with an attention mechanism to adaptively recalibrate point-wise feature responses by explicitly modeling interdependencies between different neuron units. We also introduce a non-linear face reconstruction representation as a guidance of latent space to obtain more accurate deformation, which helps solve the geometry-related deformation and is good for generalization across subjects. Huber and HSIC (Hilbert-Schmidt Independence Criterion) constraints are adopted to promote the robustness of our model and to better exploit the non-linear and high-order correlations. Experimental results on the public dataset and real scanned dataset validate the superiority of our proposed GDPnet compared with state-of-the-art model.
\end{abstract}

}

\maketitle

\IEEEdisplaynontitleabstractindextext

% \fi
\IEEEpeerreviewmaketitle

\IEEEraisesectionheading{\section{Introduction}\label{sec:introduction}}

\begin{figure*}[!t]
	\centering
	\includegraphics[width=0.8\linewidth]{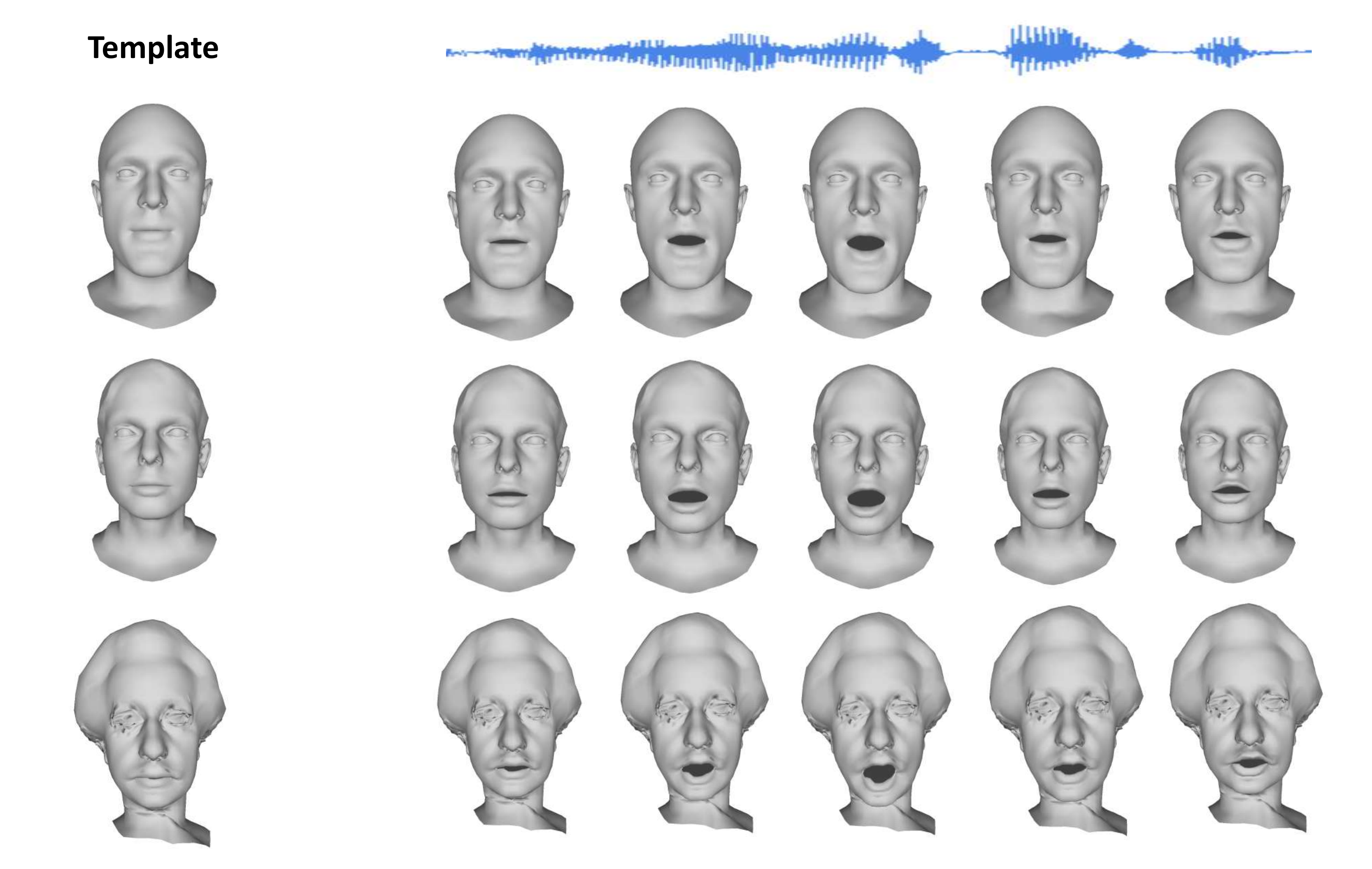}
	\caption{Our method is able to output reasonable and realistic 3D animated faces for any person in any language. Top: Actor from VOCASET \cite{Cudeiro_2019_CVPR}; Middle: Actor from D3DFACS \cite{cosker2011facs}; Bottom: Great tribute to Mr. Albert Einstein. }
	\label{fig1}
\end{figure*}

The most important approach of human communication is through speaking and making corresponding facial expressions. Understanding the correlation between speech and facial motion is highly valuable for human behavior analysis. Therefore, speech-driven facial animation has drawn much attention from both academia and industry recently, and has a wide range of applications and prospects, such as gaming, live broadcasting, virtual reality, and film production \cite{TVCG, TVCG2, Liu2015VideoaudioDR}. 3D models, as a popular and effective representation for human faces, have stronger ability to show the facial motion and understand the correlation between speech and facial motion than 2D images. However, 3D models are more complicated than images, and it is more difficult to obtain realistic 3D animation results. As shown in Figure~\ref{fig1}, our aim is to animate a 3D template model of any person according to an audio input.

Despite the great progress in speaker-specific speech-driven facial animation~\cite{Cao2005ExpressiveSF,Suwajanakorn2017SynthesizingOL,Karras2017AudiodrivenFA}, speaker-independent facial animation is still a challenging problem. Some methods  animate  unrealistic artist-designed character rigs driven by audio~\cite{DingHead,Zhou2018VisemenetAA}. Others achieve more realistic animation by combining audio and video \cite{Liu2015VideoaudioDR,Pham2018EndtoendLF}, relying on manual processes, or focusing only on the mouth \cite{Taylor2017ADL}. VOCA \cite{Cudeiro_2019_CVPR} achieves the first audio-driven speaker-independent 3D facial animation in any language using a realistic 3D scanned template. It could generate animation of different styles across a range of identities.
But there are still three challenges to achieve realistic audio-driven 3D facial animation for an arbitrary person and language:
\begin{itemize}
    \item The animated results are easily affected by both facial motion and geometry structure. Therefore, we need to consider the geometry representation of 3D models to generate more realistic animation results, in addition to relating the audio and the facial motion.
    \item The relation between audio and visual signals is complicated, and we need more effective neural networks to learn this non-linear and high-order relationship.
    \item Real signals usually contain noise and outliers, which challenge the robustness of the animation method.
\end{itemize}

In this paper, to address these challenges, we propose a geometry-guided dense perspective network (GDPnet), which consists of encoder and decoder modules. For the encoder, to ensure maximum information flow between layers in the network, we connect all layers (with matching feature-map sizes) directly with each other. For the decoder, we utilize attention mechanism to use global information to selectively emphasize informative features. Besides, we propose a geometry-guided strategy and adopt two constraints from different perspectives to achieve more robust animation. Experimental results demonstrate that the non-linear geometry representation is beneficial to the speech-driven model, and our model generalizes well to arbitrary subjects unseen during training. \emph{We will make code available to the community.}

Specifically, the main contributions of this work are summarized as follows:
\begin{itemize}
    \item We propose a dense perspective network to better model the non-linear and high-order relationship between audio and visual signals. An encoder with dense connections is designed to strengthen feature propagation and encourage the re-use of audio features, and a decoder with attention mechanism is used to better regress the final 3D facial mesh.
    \item We adopt a non-linear face representation to guide the network training, which helps to solve the geometry-related deformation and is effective for generalization across subjects.
    \item  We introduce Huber and HSIC (Hilbert-Schmidt independence criterion) constraints to promote the robustness of our model and better measure the non-linear and high-order correlations.
    \item  Our model is easy to train and fast to converge. At the same time, we achieve more accurate and realistic animation results for various persons in various languages.
\end{itemize}

\begin{figure*}[!t]
	\centering
	\includegraphics[width=0.95\linewidth]{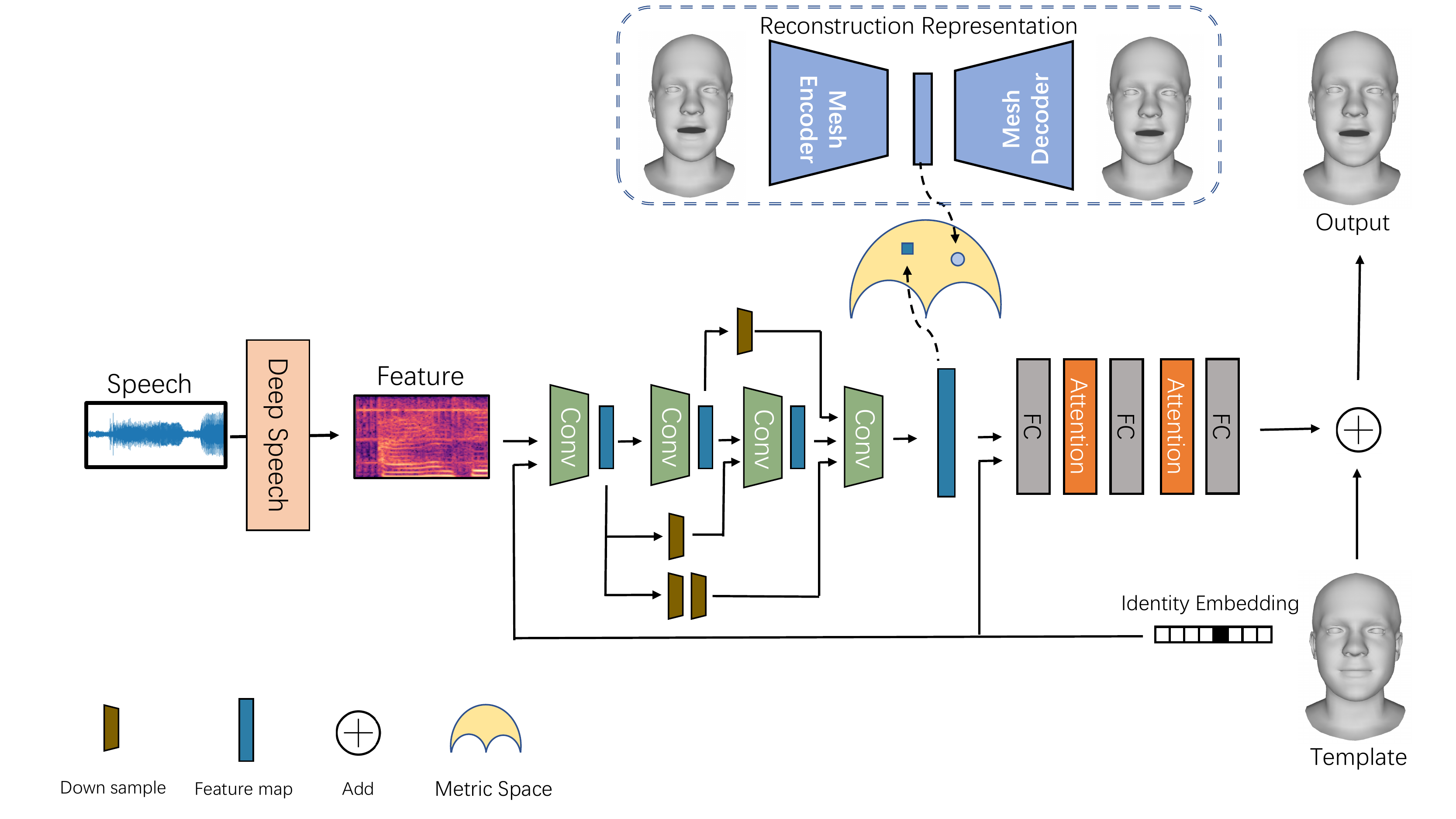}
	\caption{The architecture of our proposed geometry-guided dense perspective network. }
	\label{network}
\end{figure*}

\section{Related Work}

Despite the great progress in facial animation from images or videos \cite{kim2018deep,thies2016face2face,wu2018alive,weise2011realtime}, less attention has been paid to speech-driven facial animation, especially animating a 3D face. However, understanding the correlation between speech and facial deformation is very important for human behavior analysis and virtual reality applications. Speech-driven 3D facial animation can be categorized into two types:  speaker-dependent animation and speaker-independent animation, according to whether the method supports generalization across characters.

\subsection{Speaker-dependent Animation}
Speaker-dependent animation mainly uses a large amount of data to learn the animation ability in a specific situation. Cao \etal \cite{Cao2005ExpressiveSF} first rely on a database of high-fidelity recorded facial motions, which includes speech-related motions, but the method relies on high-quality motion capture data. Suwajanakorn \etal \cite{Suwajanakorn2017SynthesizingOL} utilize a recurrent neural network trained on millions of video frames to synthesize mouth shape from audio, but this method only focuses on learning to generate videos of President Barack Obama from his voice and stock footages. Karras \etal \cite{Karras2017AudiodrivenFA} first propose an end-to-end network for animation. Through the input of voice and specific emotion embedding, it could output the 3D vertex positions of a fixed-topology mesh that corresponds to the center of the audio window. Besides, it could produce expressive 3D facial motion from audio in real time and with low latency. However, this kind of animation methods has limited practical applications due to its inconvenience for generalization across characters.

\subsection{Speaker-independent Animation}
Many works focus on the facial animation of artist-designed character rigs \cite{kakumanu2001speech,hong2002real,salvi2009synface,taylor2012dynamic,DingHead,edwards2016jali,taylor2016audio,Taylor2017ADL,Zhou2018VisemenetAA}.
Liu \etal \cite{Liu2015VideoaudioDR} first propose a speaker-independent method based on a Kinect sensor with video and audio input for 3D facial animation, which reconstructs 3D facial expressions and 3D mouth shapes from color and depth input with a multi-linear model and adopts a deep network to extract phoneme state posterior probabilities from the audio. However, this method relies on a lot of pre-processing and inefficient search methods.
Taylor \etal \cite{Taylor2017ADL} propose a simple and effective deep learning approach for speech-driven facial animation using a sliding window predictor to learn arbitrary non-linear mappings from phoneme label input sequences to mouth movements. Pham \etal \cite{Pham2017SpeechDriven3F} propose a regression framework based on a long short-term memory (LSTM) recurrent neural network to estimate rotation and activation parameters of a 3D blendshape face model. Based on this work, they \cite{Pham2018EndtoendLF} further employ convolutional neural networks to learn meaningful acoustic feature representations, but their method also needs the recurrent layer to process the information of time series. Zhou \etal \cite{Zhou2018VisemenetAA} propose a three-stage network using hand-engineered audio features to regress the cartoon human. However, the animated face is not a realistic scanned face.
Cudeiro \etal \cite{Cudeiro_2019_CVPR} first provide a self-captured multi-subject 4D face dataset and propose a generic speech-driven 3D facial animation framework that works across a range of identities. However, none of these methods take into account the influence of geometry representation on speech-driven 3D facial animation.

In this paper, we propose a speaker-independent speech-driven 3D facial animation method by designing a geometry-guided dense perspective network. The introduced non-linear geometry representation and two constraints from different perspectives are very beneficial to achieve realistic and robust animation.

\section{Geometry-guided Dense Perspective Network}

Figure \ref{network} shows the architecture of our geometry-guided dense perspective network (GDPnet). First of all, we extract speech features using DeepSpeech \cite{Hannun2014DeepSS} and embed the identity information to one-hot embedding. After concatenating the two kinds of information, the encoder maps it to the latent low-dimensional representation. The purpose of the decoder is to map the hidden representation to a high-dimensional space of 3D vertex displacements, and the final output mesh is obtained by adding the displacements to the template.

\subsection{Problem Definition}
Suppose we have three types of data $\{(\mathbf{p}, \mathbf{x}_i, \mathbf{y}_i)\}^F_{i=1}$. Here, the index $i$ refers to a specific frame, and $F$ is the total number of frames.
$\mathbf{x}_{i} \in \mathbb{R}^{W \times D}$ is the speech feature window centered at the $i$th frame generated by DeepSpeech \cite{Hannun2014DeepSS}, where $D$ is the number of phonemes in the alphabet plus an extra one for a blank label and $W$ is the window size. $\mathbf{p} \in \mathbb{R}^{N \times 3}$ denotes the corresponding template mesh, reflecting the subject-specific geometry, and $N$ is the number of vertices of the mesh. $\mathbf{y}_{i} \in \mathbb{R}^{N \times 3}$ denotes the ground truth for facial animation at each frame. At last, let $\hat{\mathbf{y}}_{i} \in \mathbb{R}^{N \times 3}$ denotes the output of our GDPnet model for the input $\mathbf{x}_i$ with template $\mathbf{p}$.

\subsection{Model}
Our GDPnet model consists of an encoder and a decoder, as shown in Figure~\ref{network}. The input of the encoder is a DeepSpeech feature of the audio and specific identity information. In order to effectively express different subjects, we encode the identity information as one-hot embedding so as to control different speaking styles. In particular, the dimension of identity embedding is equal to the number of subjects in the training set.
During inference, changing the identity embedding alters the output speaking style.
\begin{figure*}[!t]
	\centering
	\includegraphics[width=0.9\linewidth]{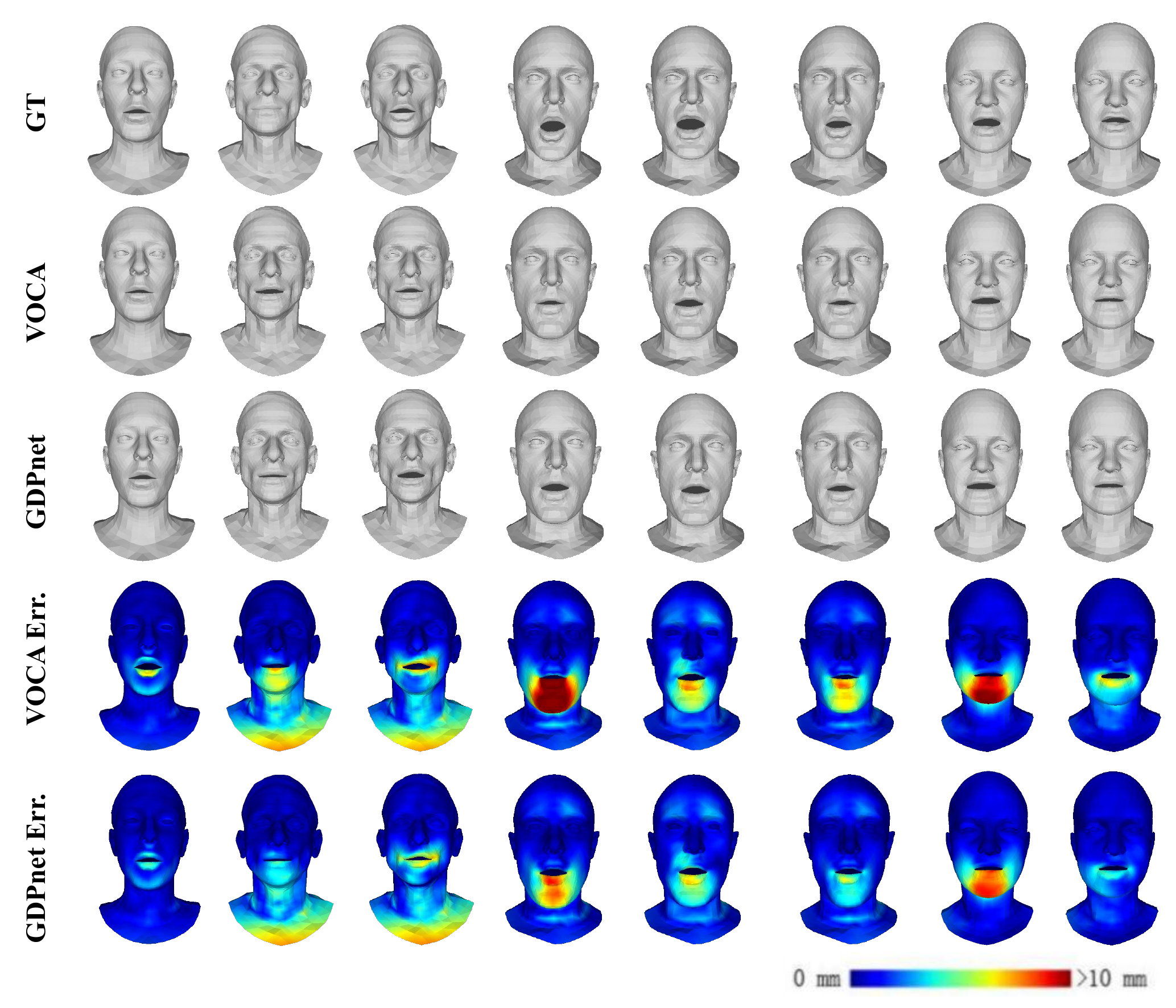}
	\caption{
    {Qualitative evaluation results for clean audio inputs.}}
	\label{fig:result}
\end{figure*}
\subsubsection{Encoder}
The purpose of the encoder is to map speech features to latent representations.
Similar to VOCA \cite{Cudeiro_2019_CVPR}, to learn temporal features and reduce the dimensionality of the input, we stack four convolutional layers for the encoder.

The problem of simply stacking the convolutional layers is that the information in the shallow layer can be easily lost \cite{Huang2016DenselyCC}.
We believe that both shallow and deep features are important, and hence we need an effective way to combine the features in the shallow layer and the deep layer. This encourages feature reuse throughout the network, and leads to more compact models. To further improve the information flow between layers, we utilize a dense connectivity pattern. Consequently, the $i$th layer receives the feature maps of all preceding layers, $\mathbf{x}_{0}, \dots, \mathbf{x}_{\ell-1}$, as input:
\begin{equation}
\mathbf{x}_{\ell}=H_{\ell}\left(\left[\mathbf{x}_{0}, \mathbf{x}_{1}, \ldots, \mathbf{x}_{\ell-1}\right]\right),
\end{equation}
where $\left[\mathbf{x}_{0}, \mathbf{x}_{1}, \dots, \mathbf{x}_{\ell-1}\right]$ refers to the concatenation of the
feature maps produced in layers $0, \ldots, \ell-1$, and $H_{\ell}$ is a composite function of two operations:
convolution (Conv) with 3 $\times$ 1 filter size and 2 $\times$ 1 stride, followed by a rectified linear unit (ReLU) \cite{Glorot2011DeepSR}.

We follow common practice and double the number of filters (feature maps) after each convolutional layer. Applying the concatenation operation in dense connections directly would be infeasible as the sizes of feature maps are different. Therefore, we introduce $2\times1$ pooling layers in the feature map dimension to reduce the number of feature maps before concatenation (indicated as the Down Sample layer in Figure~\ref{network}).
As a direct consequence of the input concatenation, the feature maps learned by any layers can be accessed easily. Benefiting from the dense connection structure, we can reuse features effectively, which makes the encoder learn more specific and richer latent representations.

\begin{figure*}[!t]
	\centering
	\includegraphics[width=0.9\linewidth]{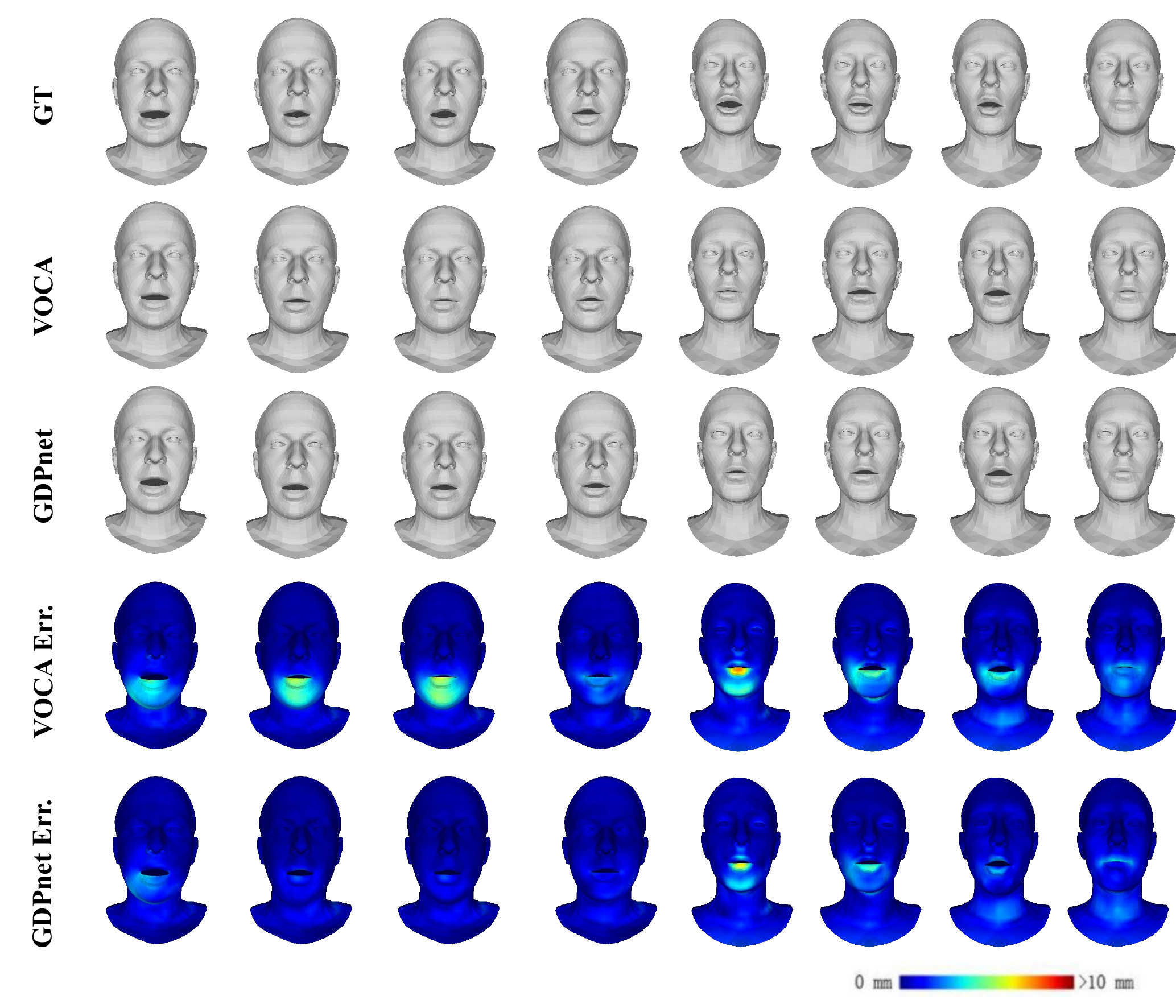}
	\caption{
    {Qualitative evaluation results for noisy audio inputs.}}
	\label{fig:noise_mse}
\end{figure*}

\subsubsection{Decoder}
The decoder maps the latent representation to a high-dimensional space of 3D vertex displacements, and the final output mesh is obtained by adding the displacements to the template vertex positions. To achieve this, we stack two fully connected layers with $\tanh$ activation function.

Inspired by the attention mechanism in image classification \cite{Hu2017SqueezeandExcitationN}, we add attention mechanism to perform feature recalibration. In this way, the network learns to use global information to selectively emphasize informative features and suppress less useful ones.
Let $\mathbf{x}_{\ell} \in \mathbb{R}^{{C} \times 1}$ denote the input of the attention layer, where $C$ is the number of feature maps,
and the attention value $\mathbf{a}_{\ell}$ can be calculated by
\begin{equation}
\mathbf{a}_{\ell} = \sigma\left(\mathbf{W}_{2} \delta\left(\mathbf{W}_{1} \mathbf{x}_{\ell}\right)\right),
\end{equation}
where $\sigma$ refers to the ReLU function and $\delta$ refers to the sigmoid function.
$\mathbf{W}_{1} \in \mathbb{R}^{\frac{C}{2} \times C}$ and $\mathbf{W}_{2} \in \mathbb{R}^{C \times \frac{C}{2}}$ denote the learnable parameter weights for the attention block. The final output of the attention block is obtained by
\begin{equation}
\widetilde{\mathbf{x}}_{l} = \mathbf{x}_{l} \otimes \mathbf{a}_{\ell}.
\end{equation}
Here, $\otimes$ is element-wise multiplication.
Through the attention block, the model can adaptively select important features for the current input samples, and different inputs can generate different attention responses.

The final output layer is a fully connected layer with linear activation function, which produces $N \times 3$ output, corresponding to the 3-dimensional displacement vectors of $N$ vertices.  $N=5023$ is used in our experiments. The final mesh can be generated by adding this output to the identity template. In order to make the training more stable, the weight of this layer is initialized by 50 PCA components calculated from the vertex displacements of the training data, and the deviation is initialized by zero.

\subsection{Geometry-guided Training}
\label{sec:training}

The encoder-decoder structure described above can be regarded as a cross-modal process.
The encoder maps the speech mode to the latent representation space, while the decoder maps the latent representation space to the mesh mode.
We refer to the latent representation $r$ as a cross-modal representation, which should express the expression and deformed geometry of a certain identity. It can be exactly related to the reconstructed expression in the 3D face representation and reconstruction using autoencoders \cite{Ranjan2018Generating3F,jiang2019disentangled,li2019generating}. The encoder encodes the input face mesh into a latent representation $\hat{r}$, and the decoder
decodes the latent representation into a reconstructed 3D mesh. In this paper, we use an MGCN (Multi-column Graph Convolutional Network) \cite{li2019generating} to extract geometry representation for each training mesh due to its ability to extract non-local multi-scale features. The geometry network is an encoder-decoder architecture with multi-column graph convolutional networks to capture features of different scales and learn a better latent space representation.

During GDPnet training, we have a mesh corresponding to a frame in each audio, and the corresponding geometry representation can be obtained using autoencoders. Using this 3D geometry representation effectively constrains the cross-modal representation.
Specifically, we want the encoder output of GDPnet to be closely related to the 3D geometry representation. Therefore, we need an appropriate measurement method to measure the relationship between them.
Here we introduce two approaches of measurement: Huber \cite{Huber1964RobustEO} constraint and Hilbert-Schmidt independence criterion (HSIC) constraint.

\begin{figure*}[!t]
	\centering
	\includegraphics[width=0.9\linewidth]{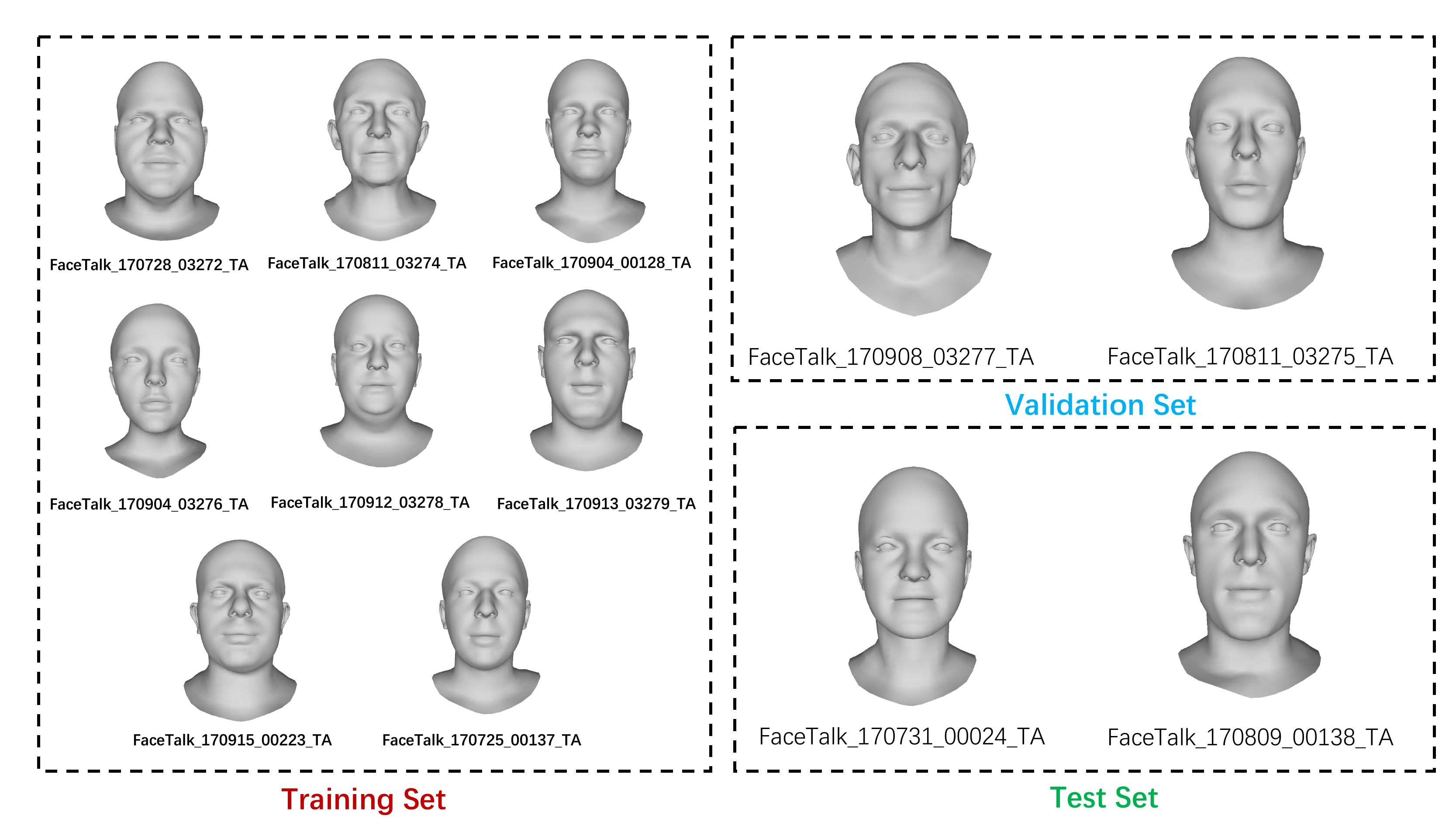}
	\caption{Specific subject names for training, validation and test.}
	\label{data}
\end{figure*}

\subsubsection{Huber Constraint} Most work uses the $\ell_2$ loss to measure the distance between two vectors, but this measurement is more easily affected by noise and outliers. $\ell_1$ loss is a better choice for robustness, but it is discontinuous and non-differentiable at position 0, leading to the difficulty for optimization. Huber loss adopts a piece-wise method to integrate the advantages of $\ell_1$ loss and $\ell_2$ loss and has been widely used in a variety of tasks.

\begin{definition}
Assuming that there are two vectors $r$ and $\hat{r}$, the Huber constraint ${L}_{\xi}$ is defined as
 \begin{equation}
 \begin{split}
 {L}_{\xi}&: \mathbb{R} \rightarrow[0,+\infty), \\
  L_{\xi}(r, \hat{r})&=\left\{\begin{array}{ll}{\frac{{r - \hat{r}}^{2}}{2}} & {\text { if }|{r - \hat{r}}| \leq \xi} \\ {\xi|r - \hat{r}|-\frac{\xi^{2}}{2}} & {\text { otherwise, }}\end{array}\right.
  \end{split}
\end{equation}
where $\xi > 0$ is the parameter that balances bias and robustness, and is set to 1.0 as default setting.
\end{definition}

The parameter $\xi$ controls the blending of $\ell_1$ and $\ell_2$ losses which can be regarded as two extremes of the Huber loss with $\xi \rightarrow \infty $ and $ \xi \rightarrow 0$, respectively.
For smaller values of $|{r - \hat{r}}|$, the loss function ${L}_{\xi}$ is  $\ell_2$ loss, and the loss function becomes $\ell_1$ loss when the magnitude of $|{r - \hat{r}}|$ exceeds $\xi$.

\subsubsection{HSIC Constraint} In addition to the distance between the two expressions, we also constrain from the perspective of correlations. If the two representations are more related, they should contain similar information. Hilbert-Schmidt independence criterion (HSIC) measures the non-linear and high-order correlations and is able to estimate the dependence between representations without explicitly estimating the joint distribution of the random variables. It has been successfully used in multi-view learning \cite{Cao2015DiversityinducedMS,Ma2019TheHB}.

Assuming that there are two variables ${R}=\left[{r}_{1}, \ldots, {r}_{i}, \ldots, {r}_{M}\right]$ and $\hat{{R}}=\left[\hat{{r}_{1}}, \ldots, \hat{{r}_{i}}, \ldots, \hat{{r}}_{M}\right]$, $M$ is the batch size.
We define a mapping $\phi({r})$ to kernel space $\mathfrak{H}$, where the inner product of two vectors is defined as $k\left({r}_{i}, {r}_{j}\right)=\left\langle\phi\left({r}_{i}\right), \phi\left({r}_{j}\right)\right\rangle$. Then, $\phi(\hat{r})$ is defined to map $\hat{r} $ to kernel space $\mathfrak{G}$. Similarly, the inner product of two vectors is defined as $k\left(\hat{r}_{i}, \hat{r}_{j}\right)=\left\langle\phi\left(\hat{r}_{i}\right), \phi\left(\hat{r}_{j}\right)\right\rangle$.
\begin{definition}
HSIC is formulated as
\begin{equation}
\begin{aligned}
 \operatorname{HSIC}\left(\mathbb{P}_{R \hat{R}}, \mathfrak{H}, \mathfrak{G}\right)  &=\left\|C_{R \hat{R}}\right\|^{2} \\ &=\mathbb{E}_{R \hat{R} R^{\prime} \hat{R}^{\prime}}\left[k_{R}\left(R, R^{\prime}\right) k_{\hat{R}^{\prime}}\left(\hat{R}, \hat{R}^{\prime}\right)\right] \\ &+\mathbb{E}_{R R^{\prime}}\left[k_{R}\left(R, R^{\prime}\right)\right] \mathbb{E}_{\hat{R}^{\prime}}\left[k_{\hat{R}}\left(\hat{R}, \hat{R}^{\prime}\right)\right] \\ &-2 \mathbb{E}_{R \hat{R}}\left[\mathbb{E}_{R^{\prime}}\left[k_{R}\left(R, R^{\prime}\right)\right]  \mathbb{E}_{\hat{R}^{\prime}}\left[k_{\hat{R}}\left(\hat{R}, \hat{R}^{\prime}\right)\right]\right], \end{aligned}
\end{equation}
where $k_{R}$ and $k_{\hat{R}}$ are kernel functions, $\mathfrak{H}$ and $\mathfrak{G}$ are the Hilbert spaces, and $E_{R\hat{R}}$ is the expectation over $R$ and $\hat{R}$.
Let $\mathcal{D}:=\left\{\left({r}_{1}, \hat{{r}}_{1}\right), \cdots,\left({r}_{m}, \hat{{r}}_{m}\right)\right\}$ drawn from $\mathbb{P}_{R \hat{R}}$.
The empirical version of HSIC is induced as:
\begin{equation}
\operatorname{HSIC}(\mathcal{D}, \mathfrak{H}, \mathfrak{G})=(N-1)^{-2} \operatorname{tr}\left({K}_{1} {H} {K}_{2} {H}\right),
\end{equation}
where $tr(\dots)$ is the trace of a square matrix. ${K_1}$ and ${K_2}$ are the Gram matrices with $k_{1, i j}=k_{1}\left({r}_{i}, {r}_{j}\right)$ and $k_{2, i j}=k_{2}\left(\hat{{r}}_{i}, \hat{{r}}_{j}\right)$.
H centers the Gram matrix which has zero mean in the feature space:
\begin{equation}
    {H}={I}_{m}-\frac{1}{m} {1}_{m} {1}_{m}^{T}.
\end{equation}
\end{definition}
Please refer to \cite{Gretton2005MeasuringSD} for more detailed proof of HSIC.

\begin{figure*}[!t]
	\centering
	\includegraphics[width=0.8\linewidth]{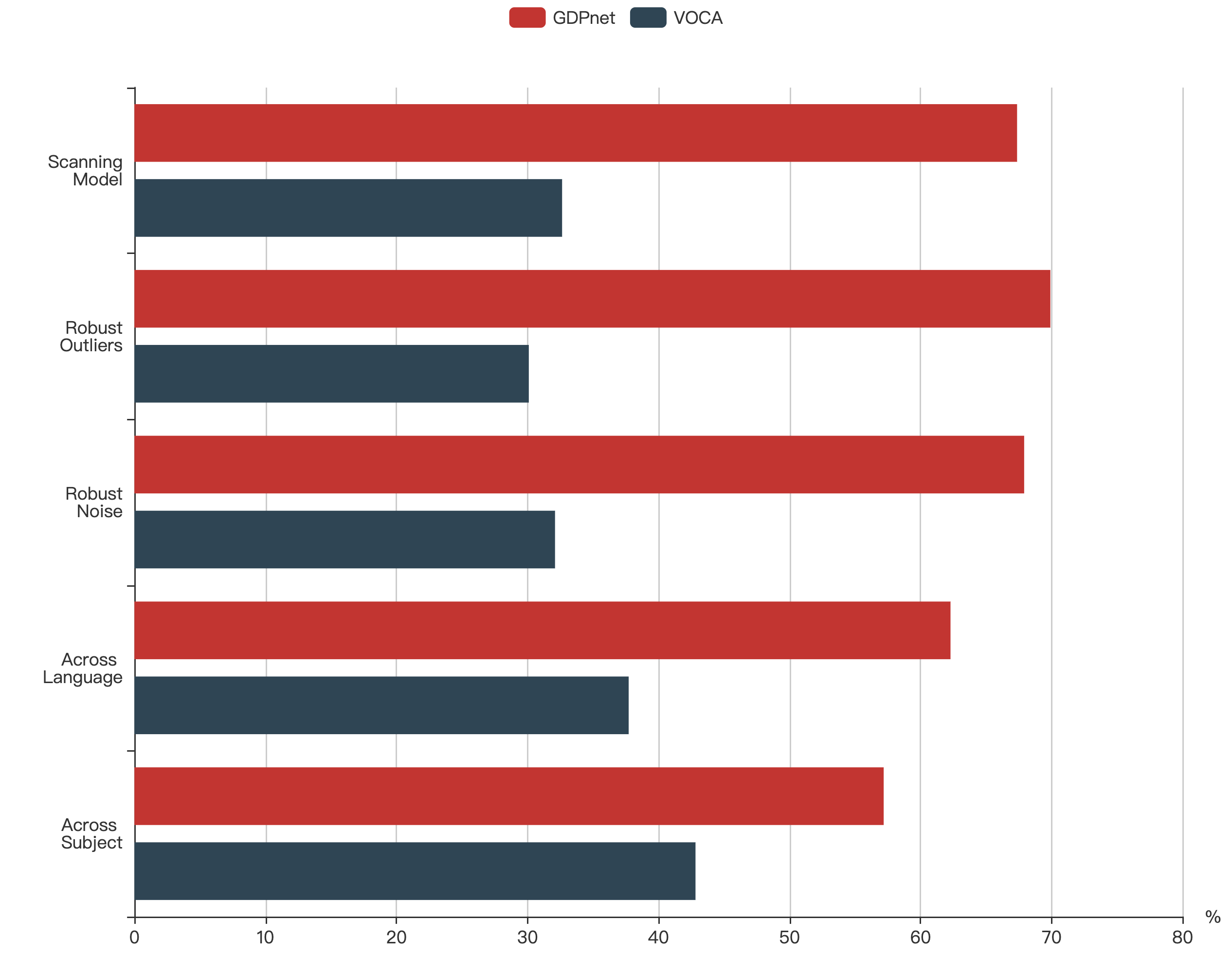}
	\caption{User study result. The bars show the percentage of users choosing VOCA\cite{Cudeiro_2019_CVPR} or ours given the same sentence for five cases. }
	\label{fig:user}
\end{figure*}

\subsection{Loss Function}
The loss of the proposed GDPnet consists of three parts, \ie, reconstruction loss, constraint loss and velocity loss:
\begin{equation}
    L = L_r + \lambda_1 L_{c} + \lambda_2 L_{v},
\end{equation}
where $\lambda_1$ and $\lambda_2$ are positive constants to balance loss terms.
The reconstruction loss $L_r$ computes the distance between the predicted output and the ground truth:
\begin{equation}
    L_{r}=\left\|\mathbf{y}_{i}-\mathbf{\hat{y}}_{i}\right\|_{F}^{2}.
\end{equation}
During the training stage, the reconstruction representation constraint $L_{c}$ could use Huber or HSIC as we discuss in Section \ref{sec:training}.
The choice of these two constraints is a trade-off, as Huber constraint has faster convergence and HSIC constraint has better performance, which will be discussed in Section \ref{sec:ablation}.
%%%YKL It is not clear whether you use one or both. If one, which one is used, and why we need to introduce both?
Besides, we have the velocity loss
\begin{equation}
    L_{v}=\left\|\left(\mathbf{y}_{i}-\mathbf{y}_{i-1}\right)-\left(\mathbf{\hat{y}}_{i}-\mathbf{\hat{y}}_{i-1}\right)\right\|_{F}^{2},
\end{equation}
to induce temporal stability, which considers the smoothness of prediction and ground truth in the sequence context.

\begin{table*}[!t]
	\small
	\caption{Performance (mm) and training time (s) of different GDPnet variants.}
	\label{ablation}
	\centering
\normalsize
\renewcommand{\arraystretch}{1.3}
	\begin{tabular}{cccccccc}
		\toprule
		{Variant} &
		{HSIC} &
		{Huber} &
		{Dense} &
		{Attention} &
		Validation &
		Test  &
		Training Time\\
%		\cmidrule{6-13}
		\midrule
		(a)&&&&& 5.861 & 7.701 & 98m58s\\
		(b)&$\checkmark$ &&&& 5.842 & 7.655 & 49m24s \\
		(c)& & $\checkmark$ & && 5.867 & 7.665  & \textbf{46m14s} \\
		(d)&$\checkmark$ &&$\checkmark$&& 5.858 & 7.628 & 49m50s \\
		(e)&$\checkmark$&&&$\checkmark$& 5.783 & 7.576 & 50m11s \\
		(f)&$\checkmark$&&$\checkmark$&$\checkmark$& \textbf{5.775} & \textbf{7.520} & 52m35s\\
		\bottomrule
	 \end{tabular}
\end{table*}

\begin{table*}[!t]
\caption{Quantitative results on VOCASET dataset (mm).}
\label{tab:result}
\centering
\normalsize
\renewcommand{\arraystretch}{1.6}
\begin{tabular}{l|ccc|ccc|c}
\toprule
\multicolumn{1}{c|}{} & \multicolumn{3}{c|}{Validation} & \multicolumn{3}{c|}{Test} & \multirow{2}{*}{Noise} \\
\cline{2-7}
& $Speaker^{val}_1$ & $Speaker^{val}_2$ & Mean & $Speaker^{test}_1$ &  $Speaker^{test}_2$ & Mean  \\
\hline
\noalign{\smallskip}
VOCA \cite{Cudeiro_2019_CVPR} & \textbf{4.073} & 7.649 & 5.861 & 9.657 & 5.844 & 7.701 & 7.890 \\
GDPnet & 4.084 & \textbf{7.467} & \textbf{5.775} & \textbf{9.377} & \textbf{5.663} & \textbf{7.520} & \textbf{7.721}\\
\bottomrule
\end{tabular}
\end{table*}

\subsection{Implementation Details}

Our GDPnet is implemented using Tensorflow \cite{Tensorflow} and trained with the Adam optimizer \cite{adam} on an NVIDIA GeForce GTX 1080 Ti GPU. We train our model for 50 epochs with a learning rate of $1e-4$ without learning rate decay. We use Adam with a momentum of 0.9, which optimizes the loss function between the output mesh and the ground-truth mesh.
The balancing weights for loss terms are set to $\lambda_1 = 0.1$ and $\lambda_2 = 10.0$, respectively.
For network architecture, we use a windows size of $W = 16$ with $D = 29$ speech features, and set the dimension of latent representation as 64.

\section{Experiments}

In this section, we first introduce the experimental setup including the dataset, training setup and the metric.
Then, we evaluate the performance of our GDPnet quantitatively and qualitatively compared with the state-of-the-art method. We also conduct a blind user study. Finally, we perform an ablation study to analyze the effects of different components of our approach.

\subsection{Experimental Setup}

\subsubsection{Dataset}
VOCASET \cite{Cudeiro_2019_CVPR} provides high-quality 3D scans with about 29 minutes of 4D scans captured at 60 fps as well as alignments of the entire head including the neck. The raw 3D head scans are registered with a sequential alignment method using the publicly available generic FLAME model \cite{FLAME}. Each registered mesh has 5023 vertices with 3D coordinates. In addition to high-quality face models, VOCASET also provides the corresponding voice data, which is very useful to train and evaluate speech-driven 3D facial animation. In total, it has 12 subjects and 480 sequences each containing a sentence spoken in English with a duration of 3-5 seconds. The sentences are taken from a diverse corpus similar to \cite{fisher1986ther}.
As we know, the posture, head rotation and other subjective information of the speaker cannot be completely judged only by voice. In order to eliminate the influence of pose and distortion on the model, we only use the unposed data for training, so that we can effectively make use of the template information to obtain more realistic animation results using the unknown voices.

\subsubsection{Training Setup}
In order to train and test effectively, we split 12 subjects into a training set, a validation set and a test set, as VOCA \cite{Cudeiro_2019_CVPR} did. Furthermore, we split the remaining subjects as 2 for validation and 2 for testing. The training set consists of all sentences of eight subjects. For the validation and test sets, 20 unique sentences are selected so that they are not shared with any other subject. The specific data division is shown in Figure \ref{data}. Note that there is no overlap between training, validation and test sets for subjects or sentences.

\begin{figure*}[!t]
	\centering
	\includegraphics[width=0.9\linewidth]{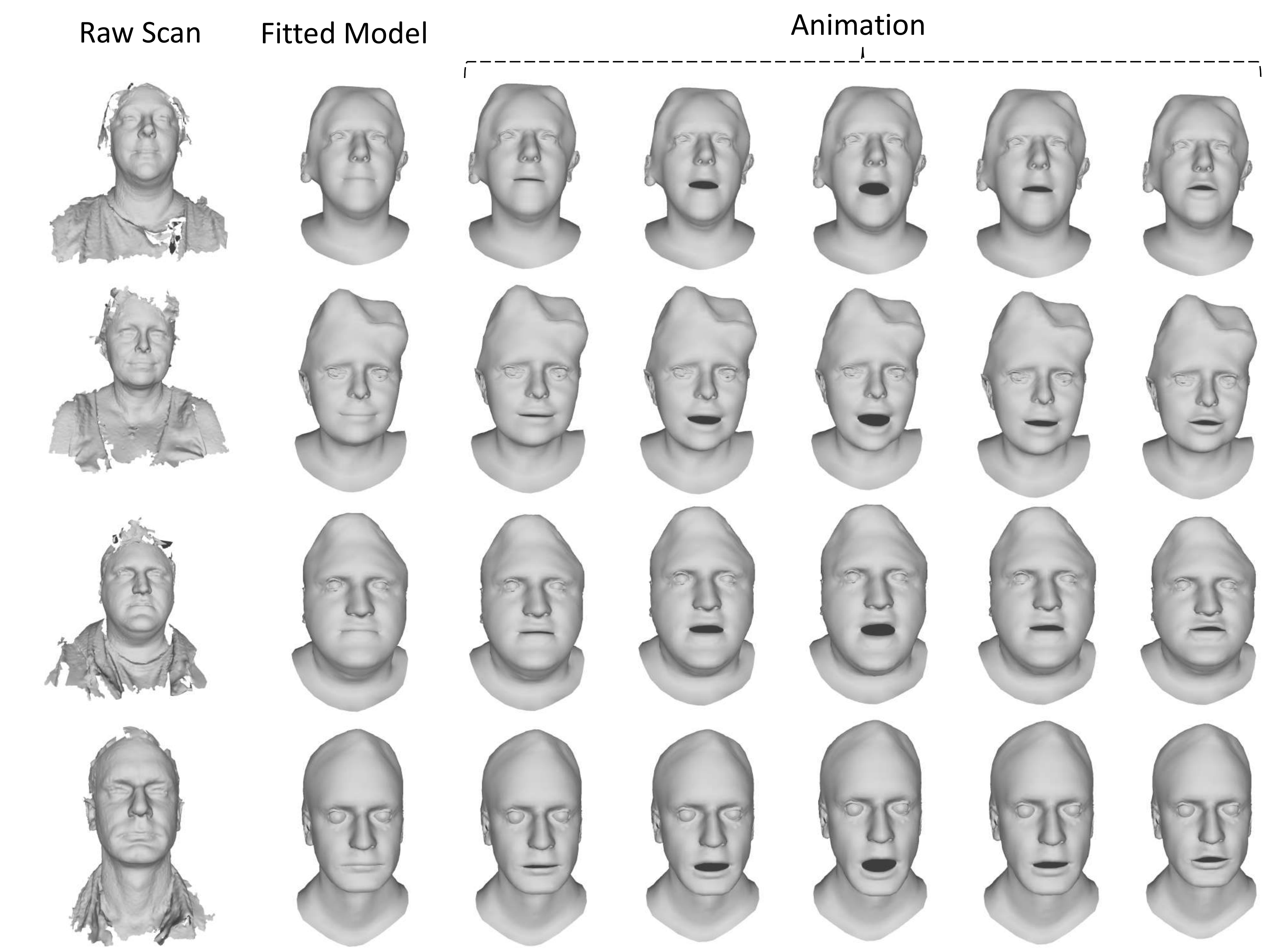}
	\caption{Our method generalizes across various scanned models from 3dMD dataset \cite{ghosh2011multiview}.}
	\label{fig:scan}
\end{figure*}

    \begin{figure*}[t]
	\centering
	\includegraphics[width=0.9\linewidth]{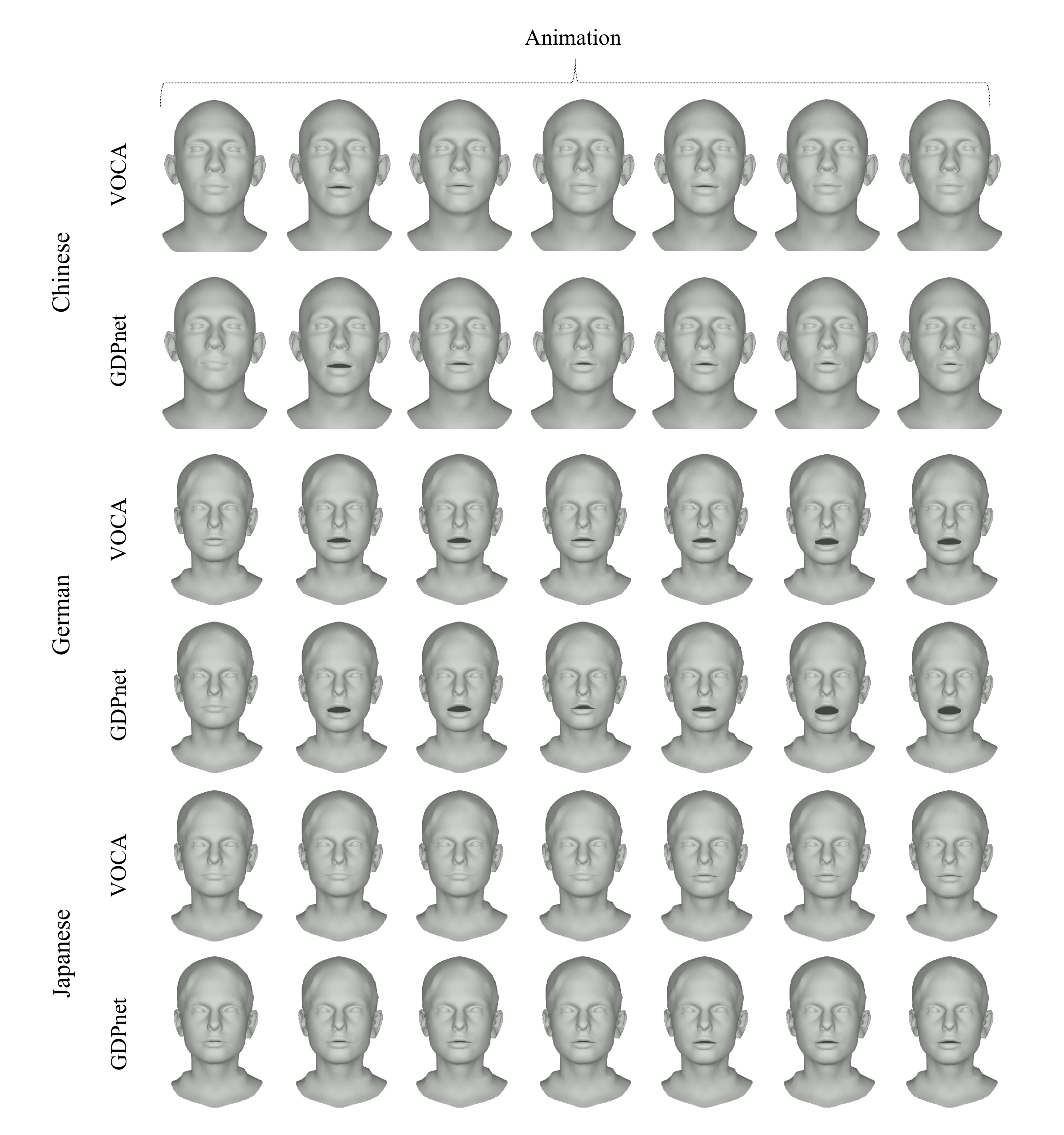}
	\caption{Our method generalizes natural and realistic animations across languages, compared with VOCA\cite{Cudeiro_2019_CVPR}.}
	\label{fig:lang}
    \end{figure*}

    \begin{figure*}[t]
	\centering
	\includegraphics[width=0.8\linewidth]{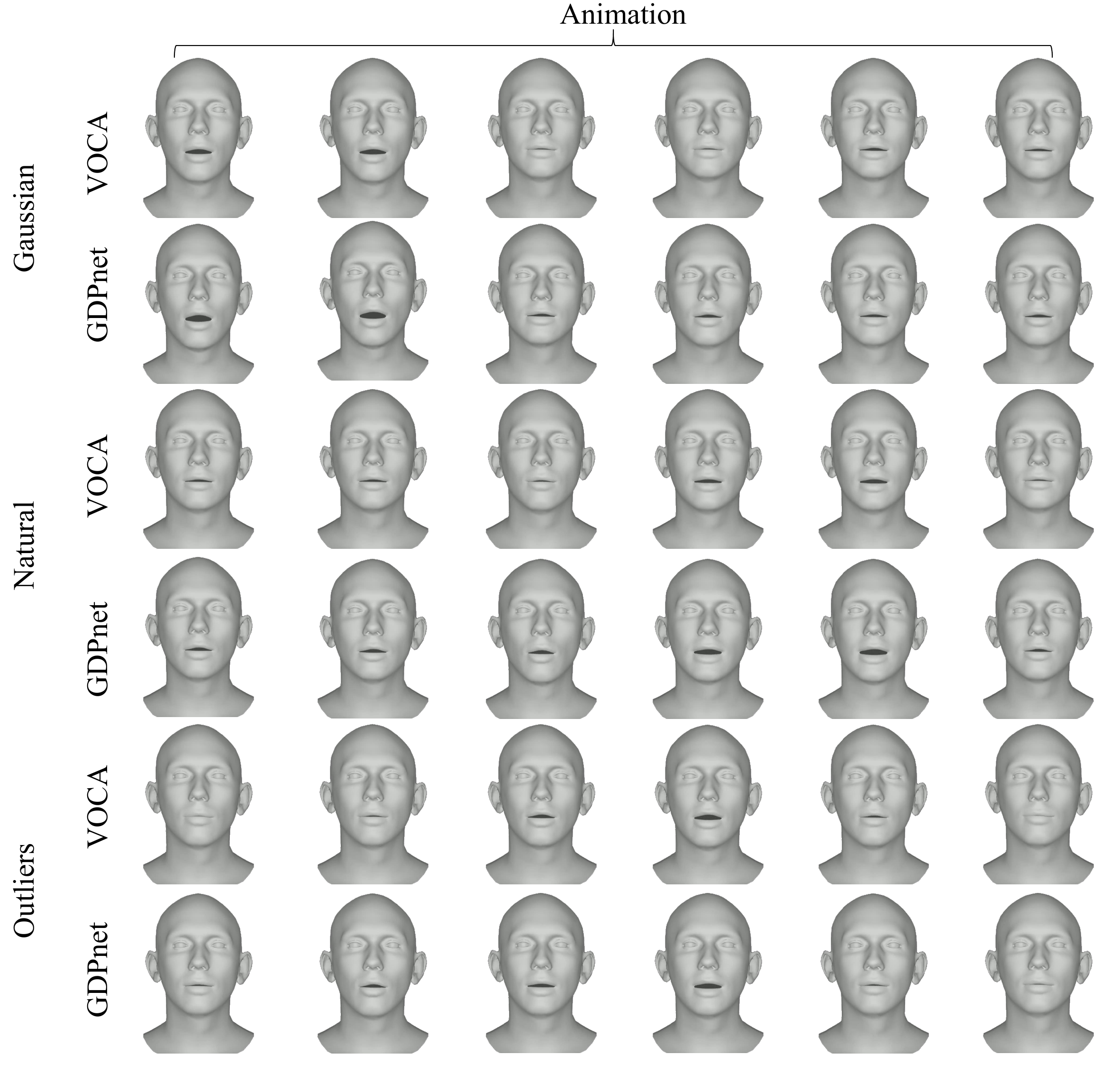}
	\caption{Our method is robust to various noise and outliers in the input audio, compared with VOCA\cite{Cudeiro_2019_CVPR}.}
	\label{fig:noise}
    \end{figure*}

\subsubsection{Metric}
To quantitatively evaluate the performances of the proposed method and the compared method, we adopt mean squared error (MSE), \ie, the average squared difference between the estimated value and the ground-truth value. Specifically, the MSE between the generated mesh $\hat{y}$ and the ground-truth mesh $y$ is defined as:

\begin{equation}
mse\left(\hat{y}, y\right)=\frac{1}{N} \sum_{i=1}^{N}\left\|{v}_{i}-\hat{v}_{i}\right\|_{2},
\end{equation}
where $v$ is a vertex of the mesh, and $N$ is the number of vertices.

\subsection{Ablation Study}
\label{sec:ablation}

Furthermore, we study the impact of different components in our GDPnet. Specifically, we analyze four key components: HSIC constraint, Huber constraint, dense connection structure in the encoder and attention mechanism.
By taking one or several components into account, we obtain six variants as follows:\\
(a) without any of the components;\\
(b) with HSIC constraint loss to leverage geometry-guided training strategy;\\
(c) with Huber constraint loss to leverage geometry-guided training strategy;\\
(d) with HSIC constraint and dense connection structure in the encoder;\\
(e) with HSIC constraint and attention mechanism in the decoder;\\
(f) with HSIC constraint, dense connect structure and attention mechanism.\\
% \begin{wrapfigure}{r}{0.3\textwidth}
%     \includegraphics[width=1\linewidth]{bad_case.pdf}
% 	\caption{One failure case using our method.}
% 	\label{fig:fail}
% \end{wrapfigure}

In Table \ref{ablation}, we compare the mean squared errors of different variants on the validation set and the test set, together with the training time. With HSIC or Huber constraint, the adoption of our geometry-guided training strategy will rapidly reduce the training convergence time of the network. The convergence speed of using Huber constraint is the fastest, because the calculation time of Huber loss is less than that of HSIC loss.
However, the performance of using Huber constraint is slightly worse than that of using HSIC constraint, since the correlation measurement of HSIC is more consistent with this task. Therefore, we use the HSIC constraint in the following comparison experiments. In summery, each module in our GDPnet can improve the performance of animation effectively, especially when using both dense connection structure and attention mechanism.

\subsection{Comparison}

In this section, we compare our method with a state-of-the-art method, VOCA \cite{Cudeiro_2019_CVPR}, quantitatively and qualitatively with a user study. VOCA \cite{Cudeiro_2019_CVPR} is the only state-of-the-art method that achieves the same goal with our work: generating realistic 3D facial animation given an audio in any language and any 3D face model.

\subsubsection{Quantitative Evaluation}

We first evaluate the quantitative results of our GDPnet method and VOCA \cite{Cudeiro_2019_CVPR} on the VOCASET dataset. For fair comparison, we use the same dataset split as VOCA \cite{Cudeiro_2019_CVPR}.
%%%YKL 'with the same random seed' is rather confusing. How is it used and which part of the evaluation is random?
%%%YKL Do we mean you use the same dataset split? If so, make it clear, as the same seed does not always translate to a fair comparison.
As presented in Table \ref{tab:result}, we calculate the mean squared error for each subject in the validation set and the test set. It can be seen that the overall performance of our model is better than VOCA, demonstrating the better generalization ability of our model.
%%%YKL You previously mentioned using either Huber or HSIC constraint. It will be useful to clarify which option is being used for the results reported in the following.
In order to more clearly formulate the different speakers in the validate and test sets, we denote the $i$th subject in the validate set as $Speaker^{val}_i$ similar to the test set.
Our GDPnet improves accuracy by $0.182mm$ on $Speaker^{val}_2$ and achieves competitive performance on $Speaker^{val}_1$ in the validation set.
It is worth noting that, in the test set, our method reduces $0.280mm$ error for $Speaker^{test}_1$ and error by $0.181mm$ for $Speaker^{test}_2$.
This proves that GDPnet is more generalized than VOCA.
Some visual results are shown in Figure \ref{fig:result}. The per-vertex errors are color-coded on the reconstructed mesh for visual inspection. Our method obtains more accurate results which are closer to the ground truths.

In order to evaluate the robustness of our method, we combine a speech signal with Gaussian noise, natural noise\footnote{From http://soundbible.com} or outliers, and use the polluted signal as the input. The averaged errors over the noisy cases are given in Table \ref{tab:result}. Our method also obtains more accurate results for the noisy cases.
Some visual results with Gaussian noisy inputs are shown in Figure \ref{fig:noise_mse}. The per-vertex errors are color-coded on the reconstructed mesh for visual inspection. More results with different noises are shown in Figure \ref{fig:noise}. These results demonstrate that our GDPnet is more robust to noise and outliers.

\subsubsection{Qualitative Evaluation and User Study}

To evaluate the generalizability of our method, we perform qualitative evaluation and perceptual evaluation with a user study, compared with the state-of-the-art method. For the user study, we show the video results of our method and VOCA \cite{Cudeiro_2019_CVPR} speaking the same sentence, and ask the users to choose the better one that is more reasonable and natural. We collect 199 answers in total, and Figure \ref{fig:user} shows the user study results.
It shows that our model gets much more votes than VOCA~\cite{Cudeiro_2019_CVPR} in the five situations.

\begin{itemize}
\item \textbf{Generalization across unseen subjects and real scanned subjects:} Our method can animate any model that has the consistent topology with the FLAME. To demonstrate the generalization capability of our method, we non-rigidly register the FLAME model against several scanned models from 3dMD dataset \cite{ghosh2011multiview}, a self-scanned model and a model of Albert Einstein downloaded from TurboSquid \footnote{https://www.turbosquid.com}. Specifically, we first manually define some 3D landmarks and fit the FLAME model to these 3D landmarks. Then, we adopt ED graph-based non-rigid deformation and per-vertex refinement to obtain a fitted mesh with geometry details. Figure \ref{fig1} shows some animation results on unseen subjects in VOCASET \cite{Cudeiro_2019_CVPR}, D3DFACS \cite{cosker2011facs}  and our fitted dataset, driven by the same audio sequence. Figure \ref{fig:scan} gives more results on our fitted dataset. Video results compared with VOCA \cite{Cudeiro_2019_CVPR} are shown in the supplemental material. Our method achieves more reasonable and realistic 3D facial animation results.\\

\item \textbf{Generalization across languages:} Although trained with speech signals in English, our model can generate animation results in any language.
 Figure \ref{fig:lang} shows some examples of our generalizations, compared with VOCA \cite{Cudeiro_2019_CVPR}. Our method achieves more reasonable and realistic results for different languages. The supplementary video gives the detailed results. \\

\item \textbf{Robustness to noise and outliers:} To demonstrate our robustness to noise and outliers, we combine a speech signal with Gaussian noise, natural noise or outliers, and use the polluted signal as the input.
   Figure \ref{fig:noise} shows a comparison between VOCA\cite{Cudeiro_2019_CVPR} and our model. Benefiting from the geometry-guided training strategy, our model not only has a faster training convergence time, but also has better robustness.
    Also, the supplementary video shows more visual results.
    \label{item:noises}

% More results can be found in the supplementary video.

\end{itemize}

\subsection{Failure Case and Discussion}

\begin{figure}[!h]
	\centering
	\includegraphics[width=1.0\linewidth]{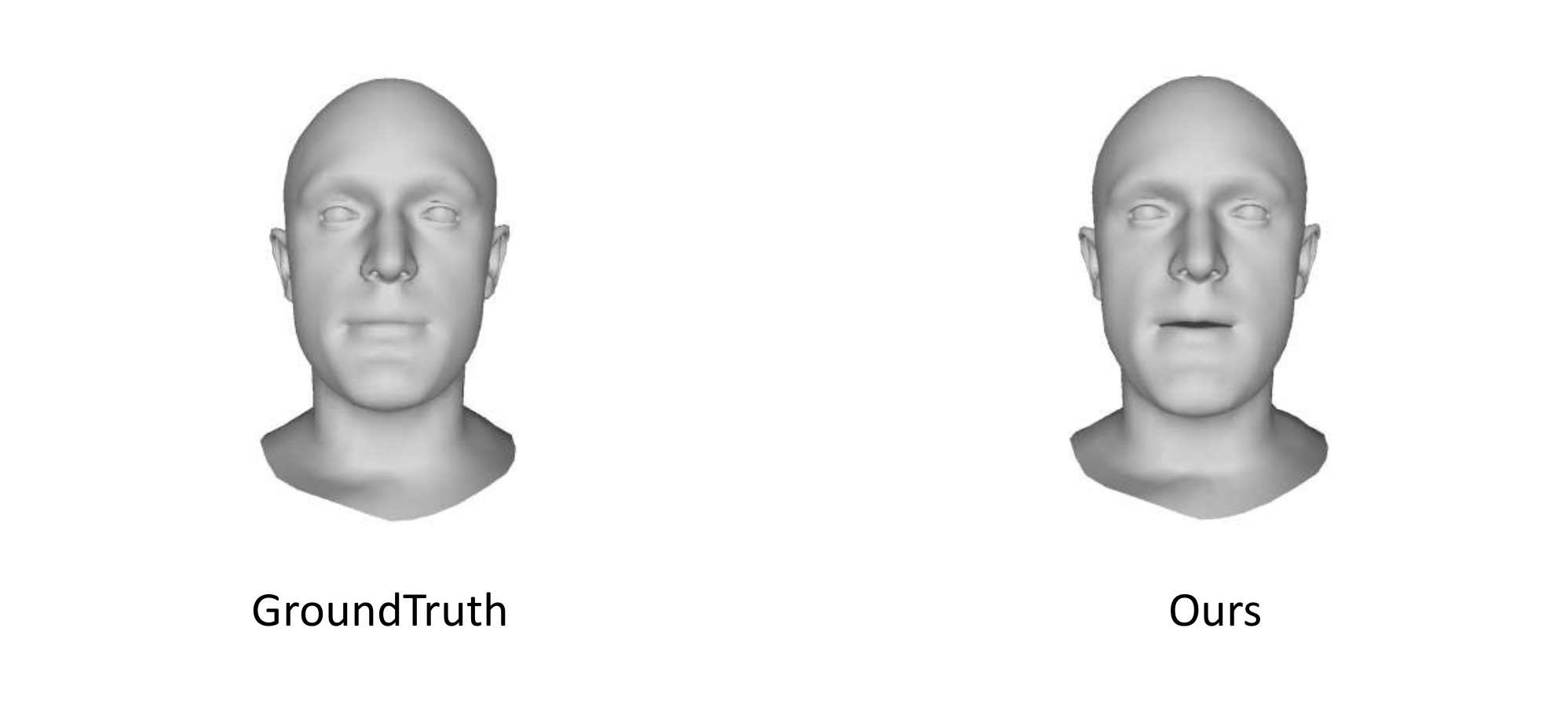}
	\caption{One failure case using our method.}
	\label{fig:fail}
\end{figure}

For the speech without face motion with the mouth fully closed, our method cannot judge the expression of speaker simply from the voice. Figure \ref{fig:fail} shows a failure case of our method. Therefore, only using audio features cannot achieve perfect 3D facial animation.
In future work, we will investigate more supervision information to assist the model in face inference, such as 2D visual information.

\section{Conclusion}

In this paper, we propose a geometry-guided dense perspective network (GDPnet) to animate a 3D template model of any person speaking the sentences in any language. We design an encoder with dense connection to strengthen feature propagation and encourage the re-usage of audio features, and a decoder with attention mechanism to better regress the final 3D facial mesh. We also propose a geometry-guided training strategy with two constrains from different perspectives to achieve more robust animation. Experimental results demonstrate that our method achieves more accurate and reasonable animation results and generalizes well to any unseen subject.

\bibliographystyle{ieee}
\bibliography{egbib}

\end{document}